\documentclass[printer]{aa}
\usepackage[english]{babel}
\usepackage{graphicx}
\usepackage{newlfont}
\usepackage{amssymb}
\usepackage{subfig}
\usepackage{commath}

\usepackage{amsmath}
\usepackage{latexsym}
\usepackage{amsthm}
\usepackage{natbib}
\usepackage{microtype}
\bibpunct{(}{)}{;}{a}{}{,} 

\usepackage[usenames,dvipsnames]{color}
\usepackage[bookmarks, colorlinks, breaklinks]{hyperref}  
\hypersetup{linkcolor=blue,citecolor=MidnightBlue,filecolor=black,urlcolor=MidnightBlue} 

\usepackage[varg]{txfonts}

\newcommand{\eqiac}[1]{Eq. \ref{#1}}
\newcommand{\seciac}[1]{Sect. \ref{#1}}
\newcommand{\figiac}[1]{Fig. \ref{#1}}

\newfont{\gwpfont}{cmssq8 scaled 1000}
\newcommand{\rexcess}{{\gwpfont REXCESS}}

\def\YX {Y_{\textrm X}}
\def\YSZ {Y_{\textrm SZ}}

\def\Mv {M_{500}}
\def\Mv {M_{500}}
\def\Rv {R_{500}}
\def\Rvyx {R_{500}^\mathrm{Y_{X}}}
\def\Mvyx {M_{500}^\mathrm{Y_{X}}}

\def\Mvhe{M_{500}^\mathrm{HE}}
\def\Rvhe{R_{500}^\mathrm{HE}}
\def\Mvher500{M^\mathrm{HE}}
\def\Mvheyx{\Mvher500\,(R<\Rvyx)}
\def\Mvsz{M^{\mathrm{SZ}}_{500}}

\def\MY {$M_{500}$--$Y_{\textrm X}$}

\def\xmm{XMM-{\it Newton}}
\def\planck{{\it Planck}}
\def\chandra{{\it Chandra}}
\usepackage{color}

\def\msol{$[10^{14} M_{\odot}]$}

\def\one{SPT$-$CL\,J2146$-$4633}
\def\two{PLCK\,G266.6$+$27.3}
\def\three{SPT$-$CL\,J2341$-$5119}
\def\four{SPT$-$CL\,J0546$-$5345}
\def\five{SPT$-$CL\,J2106$-$5844}

\begin{document}

\title{Resolving the hydrostatic mass profiles of galaxy clusters at  $z\sim1$ \\
with \xmm\ and \chandra\\}

\author{I. Bartalucci \and M. Arnaud \and G.W. Pratt \and A. M. C. Le Brun} 
\institute{IRFU, CEA, Universit\'e Paris-Saclay, F-91191 Gif Sur Yvette, France \\ 
		   Universit\'e Paris Diderot, AIM, Sorbonne Paris Cit\'e, CEA, CNRS, F-91191 Gif-sur-Yvette, France
}
\date{Submitted 13/12/17}

\abstract{We present a detailed study of the integrated total hydrostatic mass profiles of the five most massive ($\Mvsz > 5 \times 10^{14}$ M$_{\odot}$) galaxy clusters selected at $z\sim1$ via the Sunyaev-Zel'dovich effect. These objects represent an ideal laboratory to test structure formation models where the primary driver is gravity. Optimally exploiting spatially-resolved spectroscopic information from \xmm\ and \chandra\ observations, we used both parametric (forward, backward) and non-parametric methods to recover the mass profiles, finding that the results are extremely robust when density and temperature measurements are both available. Our X-ray masses at $\Rv$ are higher than the weak lensing masses obtained from the {\it Hubble Space Telescope (HST)}, with a mean ratio of $1.39^{+0.47}_{-0.35}$. This offset goes in the opposite direction to that expected in a scenario where the hydrostatic method yields a biased, underestimated, mass.
We investigated halo shape parameters such as sparsity and concentration, and compared to local X-ray selected clusters, finding hints for evolution in the central regions (or for selection effects). The total baryonic content is in agreement with the cosmic value at $\Rv$. Comparison with numerical simulations shows that the mass distribution and concentration are in line with expectations. These results illustrate the power of X-ray observations to probe the statistical properties of the gas and total mass profiles in this high mass, high-redshift regime. 
}
 \titlerunning{Total mass distribution in high-redshift galaxy clusters}
\authorrunning{Bartalucci et al.}

\maketitle

\section{Introduction}\label{sec:introduction}

In the current $\Lambda$CDM paradigm, structure formation in the Universe is driven by the gravitational collapse of the dark matter component. In this context, the form of the dark matter density profile is a sensitive test not only of the structure formation scenario, but also of the nature of the dark matter itself. In addition, it is impossible to fully comprehend the baryonic physics without first achieving a full understanding of the dominant dark matter component.

Cosmological numerical simulations uniformly predict a quasi-universal cusped dark matter density profile, whose form only depends on mass and redshift. Perhaps the best-known parameterisation of dark matter density profiles is the Navarro-Frenk-White (NFW) profile suggested by \cite{nfw}. 

This profile is flexible; in scaled coordinates (i.e. radius scaled to the virial radius) its shape is characterised by a single parameter, the concentration $c$, the ratio of the scale radius to the virial radius,$r_{s}/R_{\Delta}$\footnote{$R_{\Delta}$ is defined as the radius enclosing $\Delta$  times the critical density at the cluster redshift; $M_{\Delta}$ is the corresponding mass.}. Its normalisation, for a given concentration, is proportional  to the mass. The concentration is known to exhibit a weak dependence on mass  and redshift \citep[typically a decrease of a factor 1.5 at $z = 1$, e.g. ][]{duf08}, although the exact dependence is a matter of some debate in the literature  \citep[e.g.][]{die15}.
 
In the local ($z \lesssim 0.3$) Universe, there is now strong observational evidence for NFW-type dark and total matter density profiles with typical concentrations in line with expectations from simulations. Such evidence comes both from X-ray observations \citep[e.g.][]{pap05,vikhlinin2006,buo07}, and more recently, from gravitational lensing studies \citep[e.g.][]{mer15,oka16}. While encouraging, more work is needed to make the different observations converge, and observational biases and selection effects are still an issue  \citep[e.g. ][]{gro16}.

In contrast, constraints on distant systems, and the evolution to the present, are sparse. The recent compilation of weak and strong lensing observations of 31 clusters at $z > 0.8$ by \citet{ser13} illustrates the difficulty of obtaining firm constraints on cluster mass profiles in this redshift regime with lensing (their Fig.~1). 
Stacking the velocity data of ten clusters in the redshift range $0.87<z<1.34$, \cite{biviano2016} derived a concentration $c \equiv r_{200}/r_{-2} = 4^{+1.0}_{-0.6}$, in agreement with theoretical expectations. Perhaps the strongest constraints come from  the X-ray observations of \citet[][0.06<z<0.7]{sch08} and  \citet[][0.4<z<1.2]{amodeo2016}. 
The evolution factor of these $c$--$M$ relations, expressed as $(1+z)^\alpha$, is consistent with theoretical expectations, but with large uncertainties ($\alpha=0.71 \pm 0.52$, and $\alpha = 0.12 \pm 0.61$, respectively). 

The poor constraints at high redshift are due in part to the difficulty in detecting objects at these distances. Surveys using the Sunyaev-Zel'dovich (SZ) effect have the advantage of the redshift independent nature of the signal and the tight relation between the signal and the underlying total mass \citep{das04}. The advent of such surveys \citep{PSZ1,has13,spt2,PSZ2,hil17} has transformed the quest for high-redshift clusters. Samples taken from such surveys are thus ideal for testing the theory of the dark matter collapse and its evolution. In this context, X-ray observations, while not the most accurate for measuring the mass because of the need for the assumption of hydrostatic equilibrium (HE), can give more precise results than other methods because of their good spatial resolution and signal-to-noise ratios. A combination with theoretical modelling can give crucial insights into both the dark matter collapse and the coeval evolution of the baryons in the potential well.

Here we present a pilot study of the X-ray hydrostatic mass profiles of the five most massive  SZ-detected clusters at $z\sim 1$, where the mass is $\Mv > 5 \times 10^{14}$ M$_{\odot}$ as estimated from  their SZ signal. Initial results, obtained by optimally combining spatially and spectrally resolved \xmm\ and \chandra\ observations, concerned the evolution of gas properties, and were described in \citet[][hereafter B17]{bartalucci2016}. Here we used the same observations to probe the total mass and its spatial distribution. We discuss the various X-ray mass estimation methods used in Sect.~\ref{sec:data_sample}, and the robustness of the recovered mass distribution in Sect.~\ref{subsec:robust_profiles}. Results are compared with local systems to probe evolution in Sect.~\ref{sec:evolution} and cosmological numerical simulations in Sect.~\ref{sec:simulations}. We discuss our conclusions in Sect.~\ref{sec:conclusions}.

We adopt a flat $\Lambda$-cold dark matter cosmology with $\Omega_m = 0.3$, $\Omega_\Lambda = 0.7$, $H_{0} = 70$ km Mpc s$^{-1}$, and  $h(z) = (\Omega_m (1+z)^3 + \Omega_\Lambda)^{1/2}$ throughout. Uncertainties are given at the 68 \% confidence level ($1\sigma$). All fits were performed via $\chi^2$ minimisation. 

\section{Data sample and analysis}\label{sec:data_sample}

\subsection{Sample}
A detailed description of the sample used here, including the data reduction, is given in B17.  Briefly, the sample is drawn from the  South Pole Telescope  (SPT) and \planck\ SZ catalogues \citep{spt2, planck_xmm_plckg266}, and 
 consists of the five galaxy clusters with  the highest SZ mass proxy value\footnote{Published SPT masses are estimated `true'  mass from the SZ signal significance,  as detailed in \cite{spt2}. Masses in the \planck\  catalogue are derived iteratively from the $\YSZ$--$\Mv$ relation calibrated using hydrostatic masses from \xmm. They are not corrected for hydrostatic bias  and are on average 0.8 times smaller. In Fig.~1 of B17, and in this work, the SPT masses were renormalised by a factor of 0.8 to the \planck\ standard.} ($M_{500}^{\mathrm SZ}  \gtrsim 5 \times 10^{14}$ M$_{\odot}$)  at $z>0.9$  (see Fig.~1 of B17). 
All five objects were detected in the SPT  survey; \two\ was also independently detected in the \planck\ SZ survey. 
All five have been observed by both \xmm\ and \chandra, using the European Photon Imaging Camera (EPIC, \citealt{turner2001} and \citealt{struder2001}) and the Advanced CCD Imaging Spectrometer (ACIS, \citealt{garmire2003}), respectively. Four objects were the subject of an \xmm\ Large Programme, for which the exposure times were tuned so as to enable extraction of temperature profiles up to $R_{500}$. Shorter archival \chandra\ observations were also used. The fifth object, \two, was initially the subject of a snapshot \xmm\ observation \citep{planck_xmm_plckg266}, and was then subsequently observed in a deep \chandra\ exposure.

Dedicated pipelines, described in full in B17, were used to produce cleaned and reprocessed data products for both observatories. These pipelines apply identical background subtraction and effective area correction techniques to prepare both \xmm\ and \chandra\ data for subsequent analysis. The definition of surface brightness and temperature profile extraction regions was also identical, and point source lists were combined. 

\subsection{Analysis}

\subsubsection{Preliminaries}

Under the assumptions of spherical symmetry and HE, the integrated mass profile of a cluster is given by
\begin{equation}
M(\leq R) = -\frac{kT(r)\, r}{G\mu m_p}\left[\frac{\dif \ln{n_e(r)}}{\dif \ln{r}}+ \frac{\dif \ln{T(r)}}{\dif \ln{r}}\right],
\label{eq:mass}
\end{equation}
where $\mu=0.6 $ is the mean molecular weight in a.m.u\footnote{Any variation of the mean molecular weight with metallicity is negligible.
The typical radial or redshift dependence of metallicity in clusters \citep{mantz17} yields less than  $0.5\%$ variations on $\mu$.}, $m_H$ is the hydrogen atom mass, and $T(r)$ and $n_e(r)$  are the 3D temperature and density radial profiles, respectively.  The key observational inputs needed for this calculation are thus the radial density and temperature profiles, plus their local gradients. A complication is that these quantities are observed in projection on the sky, and thus the bin-averaged 2D annular (projected) measurements must be converted to the corresponding measurements in the 3D shell (deprojected) quantities.

A number of approaches exist in the literature for the specific case of cluster mass modelling \citep[for a review, see e.g.][and references therein]{ettori2013}. Generally speaking, one can either model the mass distribution and fit the projected (2D) quantities (backward-fitting), or deproject the observable quantities to obtain the 3D profiles and calculate the resulting integrated mass profile (forward-fitting). This deprojection in turn can either be performed either by using parametric functions or be undertaken non-parametrically. 

In the following, we chose to calculate all deprojected quantities at the emission-weighted effective radius, $r_{\mathrm w}$, assigned to each projected annulus, $i$, defined as in \citet[][]{lew03}:
\begin{equation}
r_{\mathrm w} = \left[ \left( r_{\mathrm{out_i}}^{3/2} + r_{\mathrm{in_i}}^{3/2} \right)/2 \right]^{2/3}.\label{eq:rw}
\end{equation}
Formally, $r_{\mathrm w}$ should be calculated iteratively from the density profile, but \citet{mcl99} has shown that the above equation is an excellent approximation for a wide range of density profile slopes.

\begin{figure*}[!ht]
\begin{center}
\includegraphics[width=\textwidth]{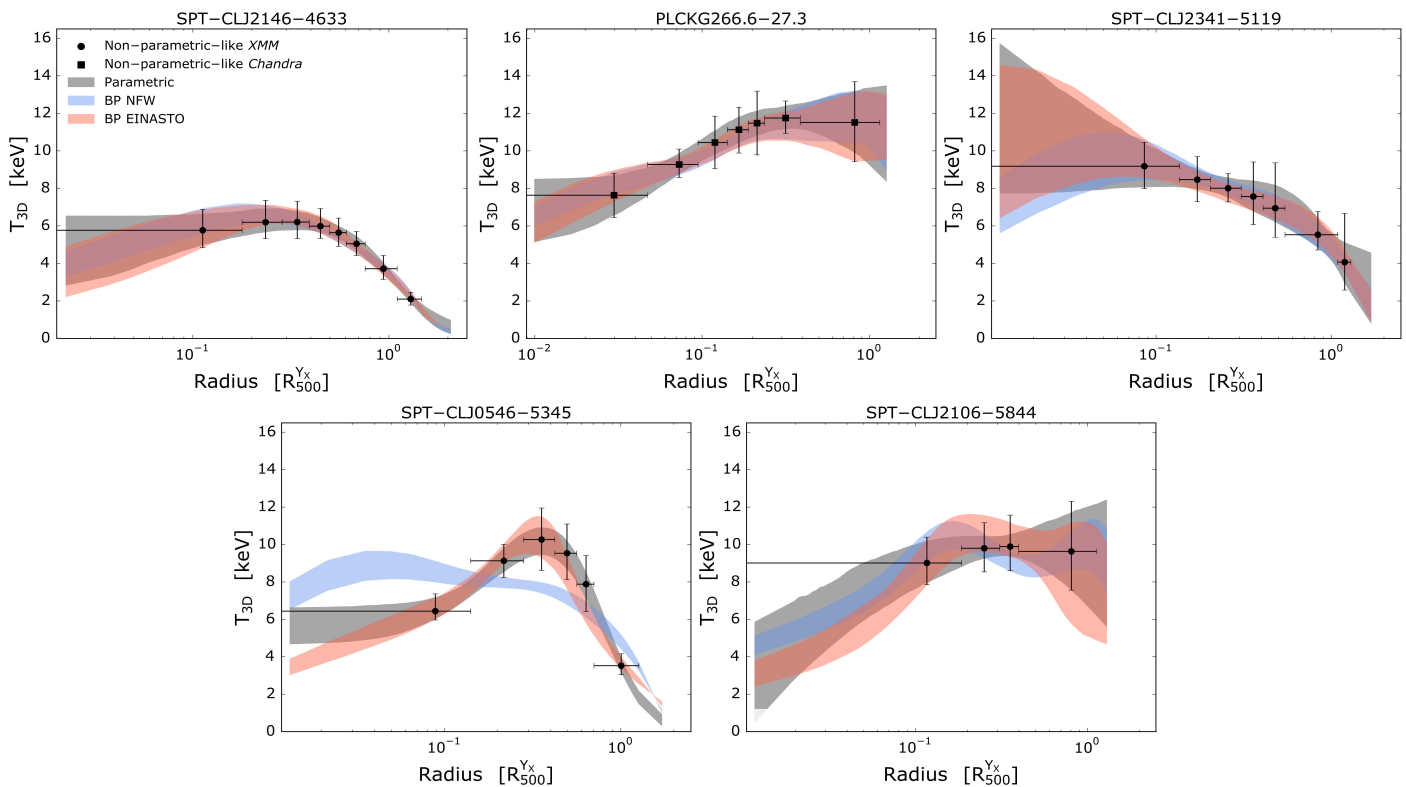} 
\end{center}
\caption{\footnotesize{3D temperature profiles of all the clusters of our sample. Radii  are scaled by $\Rvyx$. Section~\ref{sec:tprof} describes the temperature profile calculation. \textit{For each panel:} the black points represent the non-parametric-like 3D temperature profiles measured using \chandra\ and \xmm, with round and squared points, respectively. The grey shaded area represents the best-fitting 3D parametric model \citep{vikhlinin2006}. The blue and red areas represent the result of the backward fit (BP approach, see Sect.~\ref{sec:mass_prof_calc}), assuming hydrostatic equilibrium and an NFW or an Einasto mass profile, respectively. The parametric models were estimated only in the radial range covered by the density profile. The shaded regions correspond to the $68\%$ confidence level regions.}}
\label{fig:kt_profiles}
\end{figure*}

\subsection{Density and temperature profiles}\label{subsec:data_ana}

\subsubsection{Density}\label{sec:densprof}

We used the combined \xmm-\chandra\ density profiles detailed in B17, which were derived from the [0.3-2]~keV band surface brightness profiles using the regularised non-parametric deprojection technique described in \citet{croston2006}. As shown in B17, the resulting 3D (deprojected) density distributions from \xmm\ and \chandra\ agree remarkably well. 

We then fitted these profiles simultaneously with a parametric model based on that described in \citet[][see Appendix \ref{sec:parametric_models}]{vikhlinin2006}, allowing us to obtain for each object a combined density profile that fully exploits the high angular resolution of \chandra\ in the core and the large effective area of \xmm\ in the outskirts. The resulting 3D density distribution is technically a parametric profile. 
However, in view of the much better statistical quality of the density profiles (compared to that of the temperature profiles), this last parametric step does not overconstrain  the resulting mass distribution. As in B17, to avoid extrapolation, the minimum and maximum radii for the parametric models were set to match those of the measured deprojected profiles.

\four\ presents a clear substructure in its south-west sector which was not masked in B17. Since here our focus is on the measurement of integrated mass profiles, such substructures should generally be excluded from the analysis. We thus computed a new combined density profile with the substructure masked for this system. The new profile we use in this work is described in Appendix~\ref{sec:appendix_spt_ne} and is shown in \figiac{fig:spt5345ne}.

\begin{figure*}[!ht]
\begin{center}
\includegraphics[width=\textwidth]{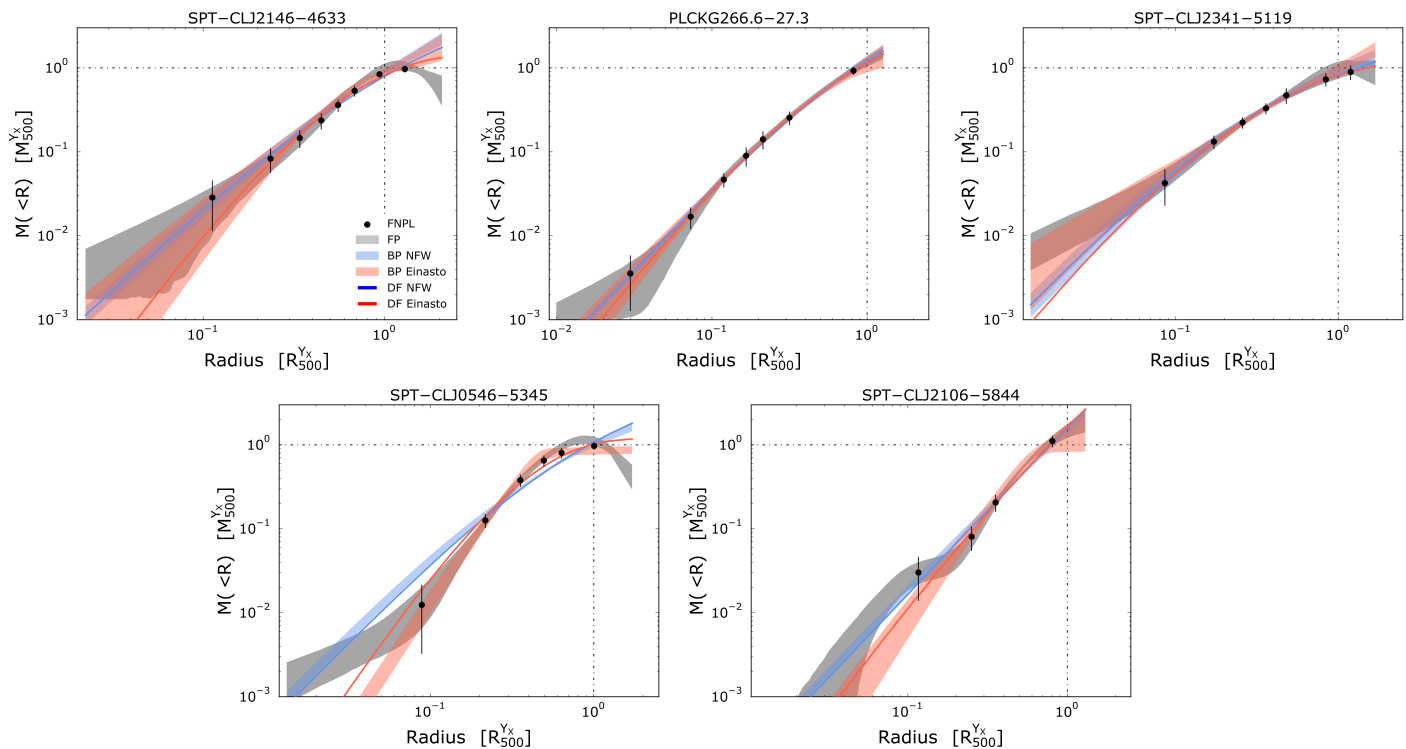} 
\end{center}
\caption{\footnotesize{Scaled mass profiles of all the clusters of our sample, derived assuming hydrostatic equilibrium (HE). All calculations used the combined  \xmm-\chandra\  density profile, and full details of the mass calculation methods are given in Sect.~\ref{sec:mass_prof_calc}. Various methods give very consistent results in the radial range with temperature information, but may diverge at small and large radius in spite of the density information. \textit{For each panel:} the black points represent the mass profiles obtained from the forward non-parametric-like method, using the HE equation and the non-parametric-like temperature profiles shown as black points in \figiac{fig:kt_profiles}. The blue and red solid lines represent the fit of  these forward non-parametric-like profiles using a NFW and an Einasto model, respectively. The grey area is the mass profile computed assuming hydrostatic equilibrium and using  the parametric temperature profiles shown with a grey area in \figiac{fig:kt_profiles}. The blue and red envelopes represent the mass profile computed using the backward method, i.e. fitting the observed temperature profile with a model derived from the HE equation and assuming a NFW and an Einasto profile, respectively, for the underlying total mass distribution. The parametric mass profiles are estimated in the wider radial range covered by the density profile. }}
\label{fig:mass_profiles}
\end{figure*}

\subsubsection{Temperature}\label{sec:tprof}

We base our 3D (deprojected) temperature profiles on those published in B17. In a first step, we extracted spectra from concentric annuli centred on the X-ray peak and determined the 2D (projected) temperature profile by measuring the temperature in each bin. 
We iteratively modified the annular binning scheme defined in B17, to ensure that  the $R_{500}^{\YX}$ fell within the outermost radius of the final annulus of each profile.

We then employed two methods to obtain the 3D temperature profile:
\begin{itemize}

\item {\it Parametric}:  We fitted a model similar to that proposed by \citet{vikhlinin2006}, reducing the number of free parameters when necessary, to the 2D profiles. This model was convolved with a response matrix to take into account projection and (for \xmm) PSF redistribution; during this convolution, the weighting scheme proposed by \citet[][see also \citealt{mazzotta2004}]{vikh_multit} was used to correct for the bias introduced by fitting isothermal models to a multi-temperature plasma. Uncertainties were computed via $1000$ Monte Carlo simulations of the projected temperature profiles.

\item {\it Non-parametric-like}: Analytical models such as those described above tend to be overconstrained, and do not reflect the fact that the temperature distribution is measured only at the points corresponding to the limited number of annuli within which spectra are extracted. To overcome these limitations we define the non-parameteric-like temperature profile by estimating the parametric model temperature at the weighted radii corresponding to the 2D annular binning scheme, and imposing the uncertainty on the annular spectral fit as a lower limit to the uncertainty in the 3D bin. 
\end{itemize}

The resulting profiles are shown in Fig.~\ref{fig:kt_profiles}, where the smooth grey envelope represents the parametric 3D temperature distribution, and the black points with errors represent the 3D non-parametric-like temperature profile.

\subsection{Mass profiles}\label{subsec:prof_mass}

\subsubsection{Mass profile calculation}\label{sec:mass_prof_calc}

Total mass profiles were determined following \eqiac{eq:mass}. To examine the robustness of the recovered profiles, we used both forward-fitting and backward-fitting methods, as we describe below.

\begin{itemize}

\item \textit{Forward non-parametric-like} (FNPL): This is our baseline mass measurement. It uses the combined 3D density profile (Sect.~\ref{sec:densprof}) and the non-parametric-like 3D temperature profile as input, and produces a mass profile estimate at each weighted radius, $r_{\mathrm w}$.  
The mass measurement and its uncertainty were calculated using a similar scheme to that first presented in \citet{pra03} and further developed in \citet{democles2010}. In this procedure, a random temperature was generated at each $r_{\mathrm w}$, and a cubic spline was used to compute the derivative. One thousand Monte Carlo simulations of this type were performed; the final mass profile and its uncertainties were then derived from the median and associated $68\%$ confidence region. The mass profiles derived from these realisations were constrained to respect the monotonic condition (i.e. $M(r+dr)>M(r)$) and to be convectively stable (i.e. $\dif  \ln{\text{T}}/ \dif \ln{n_e} < 2/3$). 

The resulting mass profiles are shown with their corresponding error bars in Fig.~\ref{fig:mass_profiles}. The relative errors are of the order of $30\%$ in the inner core, and (somewhat counterintuitively) decrease to $\sim 10-15\%$ at large radii. This effect is an intrinsic property of the typical amplitude and uncertainty on the logarithmic density and temperature gradients, and is quantified in more detail in Appendix \ref{sec:appendix_mass}.

\begin{table*}[!ht]
\caption{{\footnotesize Relevant quantities computed at fixed radii and overdensities. $M^\mathrm{DF}$ and $M^\mathrm{BP}$ are the masses computed within $\Rvyx$ using the direct fit (DF) and backward parametric (BP) methods; $c_{500}$ is derived from the DF NFW model. Radii and masses are in units of [kpc] and \msol, respectively.}}\label{tab:500_prop}
\begin{center}
\resizebox{2.\columnwidth}{!} {
\begin{tabular}{lccccccccccc}
\hline        
\hline
Cluster name        		& z & $R_{2500}^\mathrm{HE}$  		&	$\Rvhe $ &      	      $\Rvyx $      & $M_{2500}^\mathrm{HE}$ & $\Mvhe$ & $\Mvyx $ 		& $M^\mathrm{HE} (< \Rvyx)$ 	& $M^\mathrm{DF} (< \Rvyx)$ & $M^\mathrm{BP} (< \Rvyx)$ 	& $c_{500}$	\\
         &           				&              					&               		 	  &             								&      	                    &            			   &							 &								& NFW/Ein.		 &	NFW/Ein.			&	\\
\hline
SPT-CLJ2146-4633  &$0.933$ &$202_{-58}^{+33}$ &$687_{-37}^{+21}$ &$728_{-11}^{+10}$ &$ 0.34_{- 0.21}^{+ 0.19}$ &$ 2.65_{- 0.41}^{+ 0.25}$ &$ 3.15_{- 0.14}^{+ 0.13}$ &$ 2.72_{- 0.22}^{+ 0.22}$ &$ 2.50_{- 0.13}^{+ 0.12}$/$ 2.57_{- 0.14}^{+ 0.13}$ &$ 2.84_{- 0.16}^{+ 0.16}$/$ 2.86_{- 0.24}^{+ 0.24}$ & $ 1.04_{- 0.25}^{+ 0.29}$ \\
PLCKG266.6-27.3$^{a}$  &$0.972$ &$421_{-46}^{+38}$ &$1119_{-58}^{+52}$ &$993_{-14}^{+14}$ &$ 3.18_{- 0.93}^{+ 0.95}$ &$11.96_{- 1.75}^{+ 1.75}$ &$ 8.38_{- 0.36}^{+ 0.35}$ &$10.07_{- 1.08}^{+ 1.08}$ &$10.01_{- 1.13}^{+ 1.09}$/$ 9.65_{- 1.35}^{+ 1.33}$ &$10.29_{- 1.48}^{+ 1.44}$/$ 9.57_{- 2.00}^{+ 1.86}$ & $ 1.57_{- 0.32}^{+ 0.38}$ \\
SPT-CLJ2341-5119  &$1.003$ &$341_{-35}^{+38}$ &$711_{-55}^{+52}$ &$777_{-11}^{+11}$ &$ 1.76_{- 0.48}^{+ 0.65}$ &$ 3.19_{- 0.69}^{+ 0.75}$ &$ 4.16_{- 0.17}^{+ 0.17}$ &$ 3.35_{- 0.63}^{+ 0.63}$ &$ 3.44_{- 0.39}^{+ 0.39}$/$ 3.37_{- 0.42}^{+ 0.41}$ &$ 3.80_{- 0.57}^{+ 0.61}$/$ 4.02_{- 0.53}^{+ 0.58}$ & $ 4.30_{- 0.85}^{+ 1.38}$ \\
SPT-CLJ0546-5345  &$1.066$ &$389_{-39}^{+26}$ &$752_{-32}^{+28}$ &$762_{-10}^{+10}$ &$ 2.81_{- 0.76}^{+ 0.60}$ &$ 4.06_{- 0.50}^{+ 0.47}$ &$ 4.21_{- 0.16}^{+ 0.18}$ &$ 4.08_{- 0.41}^{+ 0.41}$ &$ 4.53_{- 0.34}^{+ 0.37}$/$ 4.28_{- 0.34}^{+ 0.39}$ &$ 4.30_{- 0.35}^{+ 0.37}$/$ 3.62_{- 0.38}^{+ 0.37}$ & $ 1.93_{- 0.34}^{+ 0.39}$ \\
SPT-CLJ2106-5844  &$1.132$ &$236_{-73}^{+117}$ &$1576^b$ &$880_{-19}^{+18}$ &$ 0.67_{- 0.45}^{+ 1.58}$ &$40.19^b$ &$ 7.00_{- 0.43}^{+ 0.43}$ &$10.30_{- 1.64}^{+ 1.65}$ &$11.08_{- 1.90}^{+ 1.11}$/$11.53_{- 2.39}^{+ 2.39}$ &$10.63_{- 1.01}^{+ 0.85}$/$ 8.44_{- 3.27}^{+ 3.87}$ & $ 0.01^b$ \\

\hline
\end{tabular}
}
\end{center}
\footnotesize{Notes: $^{(a)}$ SPT name: SPT-CLJ0615-5746. $^{(b)}$The $\Rvhe$, $\Mvhe$ and the $c_{500}$ values were calculated performing an extrapolation (see text for details). For this reason, these values were not used for quantitative analysis, and the errors are not reported. }
\end{table*}

\item \textit{Forward parametric} (FP): Here the fully parametric 3D density and temperature profiles were used to compute the total mass distribution on the radial grid of the combined density profile. Uncertainties were calculated using $1000$ Monte Carlo realisations, and we did not impose any condition on the resulting mass profiles. The grey shaded areas in Fig.~\ref{fig:mass_profiles} correspond to the $68\%$ dispersion envelopes.

This method may lead to non-physical results, as can be seen at large radii in \one\ and \four, where the cumulative total mass profiles start decreasing. For this reason, we do not compute a median profile and we do not use these results  to perform quantitative analyses. However, these profiles retain the maximum amount of information on the intrinsic dispersion, allowing us to explore the dispersion related to density and temperature measurement errors. Additionally, these mass profiles are estimated on the finer radial grid and wider radial range of the density profiles and so  they can be used to qualitatively investigate the behaviour in the cluster core and outskirt regions. We note that we did not extrapolate the parametric model of the density profiles, i.e. we did not attempt to estimate masses  in regions where there are no observational constraints.

\item \textit{Backward parametric} (BP): Here we assumed that the total mass distribution could be described by an NFW \citep{nfw} or Einasto \citep{ein65,nfw04} distribution, and inverted \eqiac{eq:mass}, taking into account the 3D density profile, to obtain  the corresponding 3D temperature profile. This was then projected and convolved with the instrument response and PSF, and fitted to the 2D temperature profile. Uncertainties were estimated through a Monte Carlo randomisation procedure using 1000 realisations. The resulting temperature and mass profiles are shown in Figs.~\ref{fig:kt_profiles} and~\ref{fig:mass_profiles}. The analysis was again restricted to the radial range covered by the density profile.

\item {\it Direct fit} (DF): We also directly fitted the FNPL mass profiles using the NFW and Einasto functional forms.  The resulting best fits, computed on the combined density profile radial grid, are shown in Fig.~\ref{fig:mass_profiles}. The corresponding uncertainties were estimated by repeating the fitting procedure on $1\,000$ Monte Carlo realisations of the FNPL mass profile. The NFW fit concentrations at $R_{500}$, $c_{500} \equiv R_{500}/r_{s}$ where $r_{s}$ is the scale radius, are given in Table~\ref{tab:500_prop}. 

\end{itemize}

\subsubsection{Determination of mass at fixed radius and density contrast}\label{subsec:delta_mass}

The value of $\Mvyx$  (and consequently $\Rvyx$) was determined iteratively using the \MY\ relation, as calibrated in \cite{arnaud2010}, assuming self-similar evolution. Here $Y_{ X}$ is defined as the product of the gas mass computed at $\Rvyx$ and the temperature measured in the $[0.15-0.75]\Rvyx$ region \citep{krav2006}. As the radial density bin widths used here differ from those used in B17, as described above in \seciac{subsec:data_ana}, the gas mass profiles and the quantities based on \MY were updated. For this reason, the values  in Table~\ref{tab:500_prop}  differ slightly ($\sim 1\%$) from those published in Table~2 of B17.

We determined the FNPL masses at density contrasts $\Delta=[2500,500]$, namely $M_{2500}^\mathrm{HE}$ and $\Mvhe$, at  radii $R_{2500}^\mathrm{HE}$ and $\Rvhe$, respectively. We also interpolated all the mass profiles (except FP) described in Sect.~\ref{subsec:prof_mass}, at $\Rvyx$. These are referred to as $M^\mathrm{Method}\,(R<\Rvyx)$ in the following text and figures. Radii and the corresponding masses are given in Table~\ref{tab:500_prop}.

The $\Mvhe$ of \five\ reaches a non-physical value of $\sim 40 \times 10^{14} M_{\odot}$. 
The $\Rvyx$ is at the outer edge of the last temperature bin, so extrapolation is required. As the mass profile of this object is very steep, the radius at which $\Delta=500$ is boosted, and the corresponding mass reaches non-realistic values.  The resulting $\Mvhe$ and $\Rvhe$ estimates are provided in Table~\ref{tab:500_prop}, although they are not used for any quantitative analysis. The DF NFW yields more reasonable $\Mv$ estimates, although they are poorly constrained. 
The best-fit $c_{500}$ value is equal to the minimum value allowed by the fit ($c_{500}=0.01$), corresponding to the quasi-power law behaviour of the mass profile, and yields  an $\Mv$ that is significantly greater than  $\Mvyx$.  A more conservative lower value of $c_{500}=1$ forces the curve to be higher in the core and the fit is then driven by the third point ($R \sim R_{2500}$) because of its small relative error. This analysis yielded a $\sim 5$ times higher $\chi^2$ and a value of $M_{500}=7.6 \pm 2.1 \times 10^{14} M_{\odot}$, now in agreement with $\Mvyx$.  This result must simply be considered as an NFW extrapolation, with priors on $c_{500}$,  of the well-determined mass at $R_{2500}$.

Two objects from our sample, \four\ and \five, were also analysed by \citet{amodeo2016} using \chandra\ only datasets. The authors estimated $M_{200}$ and $c_{200}$ using the BP approach and the NFW functional form. Using the concentration and mass values published in their Table~2 to compute $M_{500}$ yields $M_{500}=4.0 \pm 2.9 \times 10^{14}M_{\odot}$ and $M_{500}=6.5 \pm 3.9 \times 10^{14}M_{\odot}$ for \four\ and \five, respectively. These are perfectly consistent with the present  BP-NFW estimates; however, our deeper observations and extended radial coverage  allowed us to better constrain the measurements, the relative errors being $\sim 5$ times smaller.

\section{Robustness of X-ray mass}\label{subsec:robust_profiles}
In this section,  we first examine  the robustness of the HE mass estimate to  the X-ray analysis method. As the HE assumption is a  known source of systematics, through the HE bias, we then compare the HE  mass to lensing mass estimates, which do not rely on this assumption. 
\subsection{Mass profile shape}

Figure \ref{fig:mass_profiles} shows the mass profiles resulting from the different mass estimation methods discussed above. 
The BP results indicate that while the NFW model is a good description in the case of relaxed objects (e.g. \two) and some perturbed systems (e.g. \three), the Einasto model is generally a better fit for our sample (as is evident from the figure, and from the $\chi^2$ value) and is more able to fit a wider range of dynamical states. This is unsurprising given the larger number of parameters in the Einasto model. Forward and backward methods also give extremely consistent results.  The limitations of the NFW model can be seen in \four, where this form is clearly a poor description of the data, leading to the BP NFW masses being somewhat different to those from other methods. 

Overall, all the mass estimation methods yield remarkably robust and consistent results within the radial range covered by the spectroscopic data, i.e. within the minimum  and maximum  effective radii of the temperature profile bins, except in cases where the underlying model is insufficiently flexible. Mass profile uncertainties are quite different between methods, however, with the FP method yielding the smallest and the FNPL method yielding the largest (or most conservative). 
This simply reflects the restrictions each method places on the possible shape of the profile.

Outside the radial range covered by the spectroscopic data, the results are most robust and agnostic to the mass estimation method when the profiles are regular and can be described by a simple model (e.g. NFW). However, when the radial sampling is poor (the profiles have few points) or when the profile is irregular (e.g. \four), estimation of the mass outside the radial range probed by the spectroscopic data is less robust and will depend strongly on the method used to measure the mass. In addition, outside the region covered by the spectroscopic data, the uncertainties  rapidly increase with the distance from effective radius of the final temperature measurement, in spite of the density information.

\begin{figure}[!t]
\begin{center}
\includegraphics[width=0.8\columnwidth]{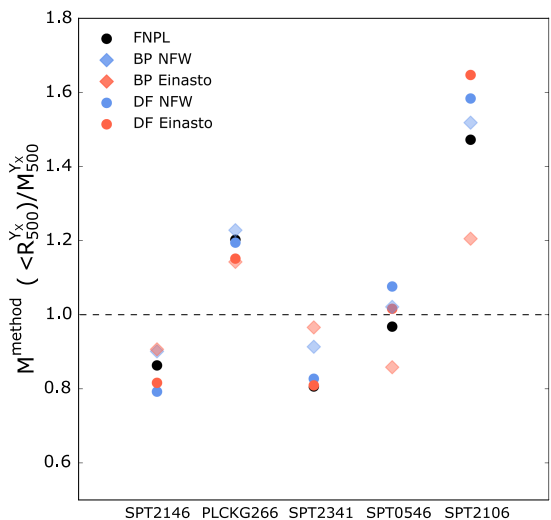} 
\end{center}
\caption{\footnotesize{Comparison of the hydrostatic mass computed at fixed radius, $\Rvyx$,  using the different methods, in units of  $\Mvyx$.  There is excellent agreement, with differences of less than  $10\%$,  when  the radius is  enclosed in the radial range covered by the spectroscopic data.} }
\label{fig:ratio_masses}
\end{figure}

\begin{figure*}[!t]
\begin{center}
\includegraphics[width=0.8\textwidth]{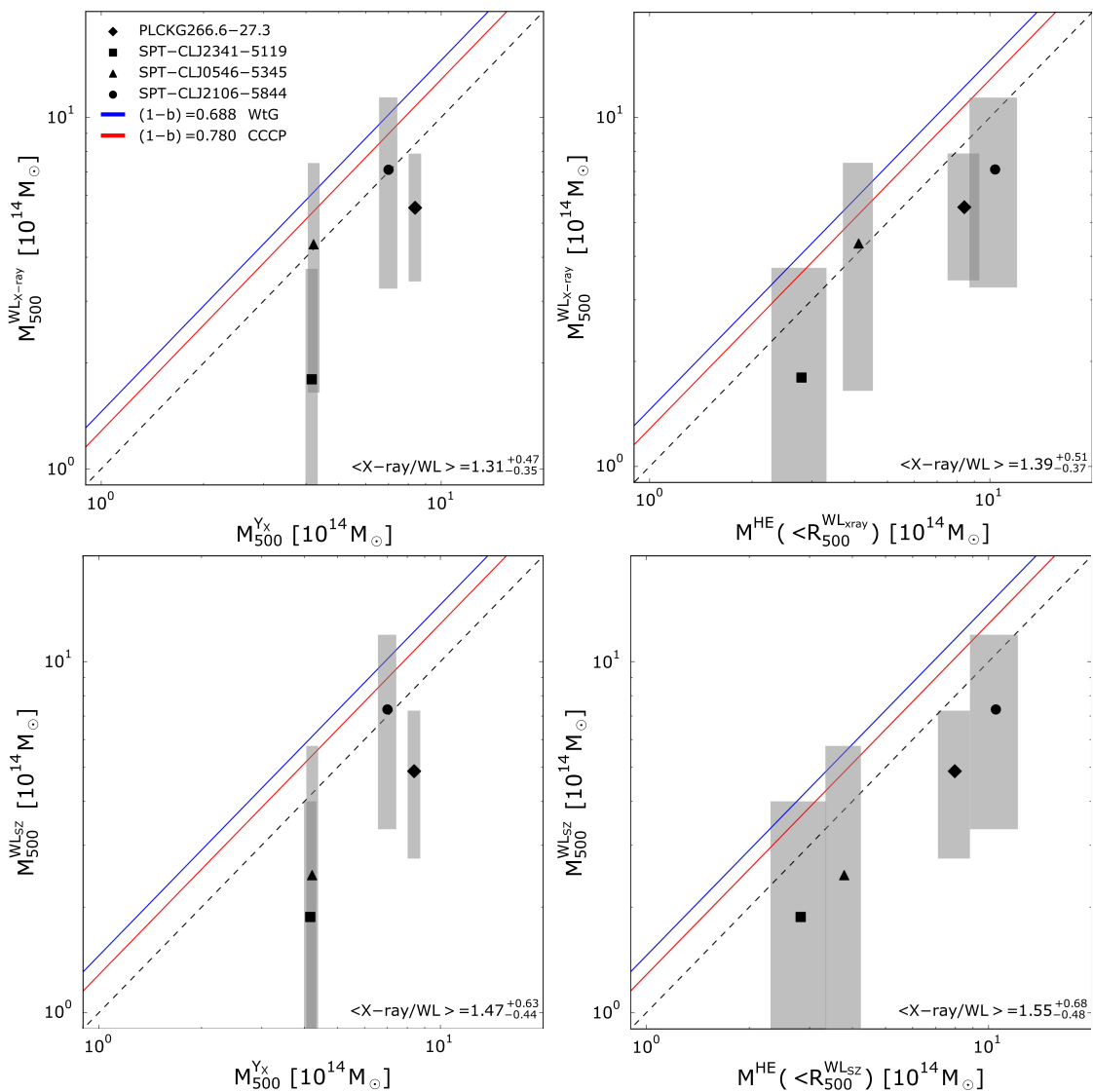} 
\end{center}
\caption{\footnotesize{Comparison between our X-ray  masses and the weak lensing masses published in \citet{tim2016}.  All estimates for a given cluster are consistent within the statistical errors. However, there is a general trend of smaller lensing mass than the HE mass, contrary to expectation. Note also the higher statistical precision of the X-ray masses. \textit{Top left panel:} comparison between $\Mvyx$ and weak lensing masses estimated at $R_{500}$, centred on the X-ray peak. The grey area for the weak lensing represents the statistical errors. The black solid bars represent the sum in quadrature of systematic and statistical errors. The blue and red lines represent the bias, $(1-b)$, between the X-ray hydrostatic and weak lensing mass as measured by Weighting the Giants (WtG, \citealt{wtg}) and by the Canadian Cluster Comparison Project (CCCP, \citealt{cccp}), respectively. To better visualise the points, we crop the lower values of SPT-CLJ2341-5119 and SPT-CLJ0546-5345, which are of the order of $\sim 10^{-1}\times 10^{14}M_{\odot}$. \textit{Top right panel:} same as the top left panel, except showing the comparison between the hydrostatic mass  computed at $\Rvyx$, $\Mvheyx$, and the weak lensing masses. \textit{Bottom left and right panels:} same as the top panels except that weak lensing masses are computed using the Sunyavez-Zeldovich (SZ) peak as the centre. The error--weighted mean ratio  and corresponding errors are reported in each panel.}}
\label{fig:lensing_xpeak}
\end{figure*}

\subsection{Mass within $\Rvyx$}
\label{subsec:M500}

\begin{figure*}[!t]
\begin{center}
\includegraphics[width=0.8\textwidth]{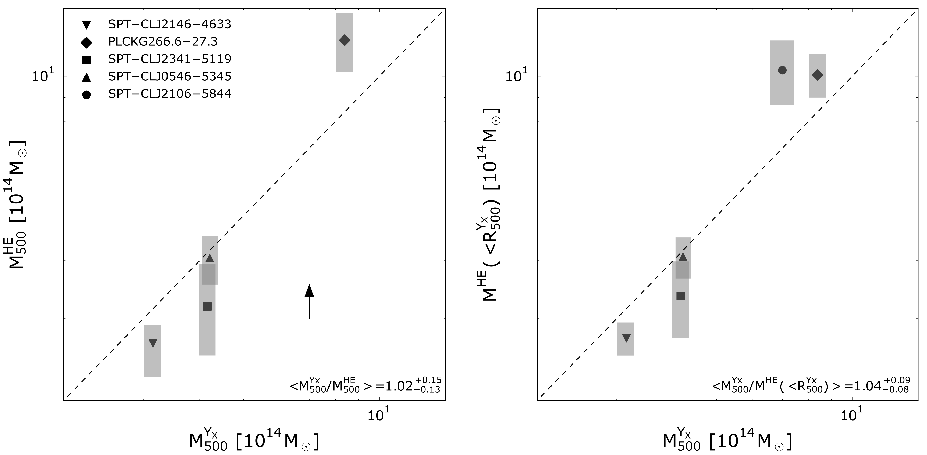} 
\end{center}
\caption{\footnotesize{\textit{Left panel}: Comparison between  $\Mvyx$ computed iteratively through the \MY\ relation and the hydrostatic mass, $\Mvhe$. The estimates are consistent within the statistical errors.  The $\Mvhe$ of \five\ is $\sim 3.5$ greater than $\Mvyx$ (see \seciac{subsec:delta_mass}). For this reason, the point is off the scale and its $\Mvyx$ is instead shown with the black arrow. The black dotted line is the $1:1$ relation. \textit{Right panel}: same as the left panel, except showing the comparison between $\Mvyx$ and $\Mvheyx$.  Error--weighted mean ratio  and corresponding errors are reported in each panel.}}
\label{fig:yx_vshy}
\end{figure*}

We now turn to the robustness of the mass determined within a fixed radius, calculated as described in \seciac{subsec:delta_mass}. Figure \ref{fig:ratio_masses} shows the ratio between the mass obtained employing the different methods, with $\Mvyx$ as a reference mass. For \one\ and \two\ the HE mass measurements are in excellent agreement, the difference being within a small percent. \three\ and \four\ present larger differences ($\sim 10\%$) according to the mass estimation method between mass estimates. Interestingly, the BP masses of \three\ are closest of all the objects to its $\Mvyx$.

\five\ is the only cluster for which all the methods yield masses greater than $\Mvyx$, by a factor of $\sim 40\%$, except if we further restrict the possible range of concentration parameters. The difference between mass estimates is also noticeably larger than for the other objects,  due to the limited radial coverage. Even  if the masses are estimated at a fixed radius, $\Rvyx$, this radius falls barely within the outermost temperature radial bin. We conclude that that in order to perform robust measurements, the  radius at which the mass is to be estimated should lie within the weighted radial range covered by the spectroscopic data.

\subsection{Comparison to weak lensing}\label{subsec:mdelta_vs_lensing}

Weak lensing mass measurements represent an additional and independent method of investigating the robustness of our mass determinations; furthermore, understanding the systematic differences between  weak lensing and X-ray masses at $z \sim 1$ is crucial for any future cosmological or physical exploitation of such samples. We compared our results with the weak lensing masses published in \citet{tim2016}, who determined  $M_{500}$ for $13$ SPT clusters observed with the \textit{Hubble Space Telescope}. Four of their objects are in common with our sample.

\citet{tim2016} give different weak lensing $\Mv$ estimates,  depending on the choice of centre (X-ray peak and SZ peak). The top left panel of \figiac{fig:lensing_xpeak} show the comparison between $\Mv$  measured using the X-ray peak as centre, $M_{500}^\mathrm{WL_{X-ray}}$, and  $\Mvyx$, as listed in Table~\ref{tab:500_prop}. Formally, there is good agreement, with  the masses for each individual cluster being consistent at $1\sigma$. However, there is a clear systematic offset in the sense that all X-ray masses are higher than the WL masses, with an error-weighted mean ratio of $\Mvyx/M_{500}^\mathrm{WL_{X-ray}}=1.31^{+0.47}_{-0.35}$. The right panel  shows the comparison with the HE masses, computed  at  $R_{500}^\mathrm{WL}$ (instead of $\Rvyx$) to avoid an artificial increase in differences due to different apertures. The difference is similar to $\Mvhe/M_{500}^\mathrm{WL_{X-ray}}=1.39^{+0.51}_{-0.37}$.  We found the same results by comparing the X-ray masses   with the weak lensing masses centred on the SZ peak, $M_{500}^\mathrm{WL_{SZ}}$, as shown in the bottom panels of \figiac{fig:lensing_xpeak}.  

This result is unexpected. The so-called `hydrostatic bias', owing to the assumption of HE, is believed to result in a net underestimate of the total mass in X-ray measurements, while lensing observations, although slightly biased, are expected to yield results that are closer to the true value. Indeed, such a trend has been found, for example in the Weighing the Giants (WtG, \citealt{wtg}) project and by the Canadian Cluster Comparison Project (CCCP, \citealt{cccp}), where the X-ray hydrostatic masses are $\sim 30\%$ and $20\%$ {\it lower} than the WL values\footnote{These works express the bias in terms of $M_{x} = (1-b)M_{500}^\mathrm{WL}$, where $M_{\mathrm X}$ is the hydrostatic X-ray mass and b is the bias between the measurements which encodes all the systematics.}, respectively. While there are only four objects in our sample, we find the opposite trend here.  With a HE-to-WL mass ratio of  $1.39^{+0.51}_{-0.37}$,  our results are  marginally consistent at $1\sigma$ with the  \citet{tim2016} results, and inconsistent with WtG at the $2\sigma$  level.

This comparison underlines  the capability and complementarity of X-ray observations with respect to optical observations, especially at these redshifts. The X-ray statistical errors are significantly smaller than the weak lensing uncertainties; furthermore, the X-ray results are remarkably robust, as we demonstrate in the previous sections. The results we find here show that while X-ray observations at high redshift are expensive and challenging, they offer a robust and precise tool which can efficiently complement measurements in other wavelengths.

\begin{figure*}[!t]
\begin{center}
\includegraphics[width=\textwidth]{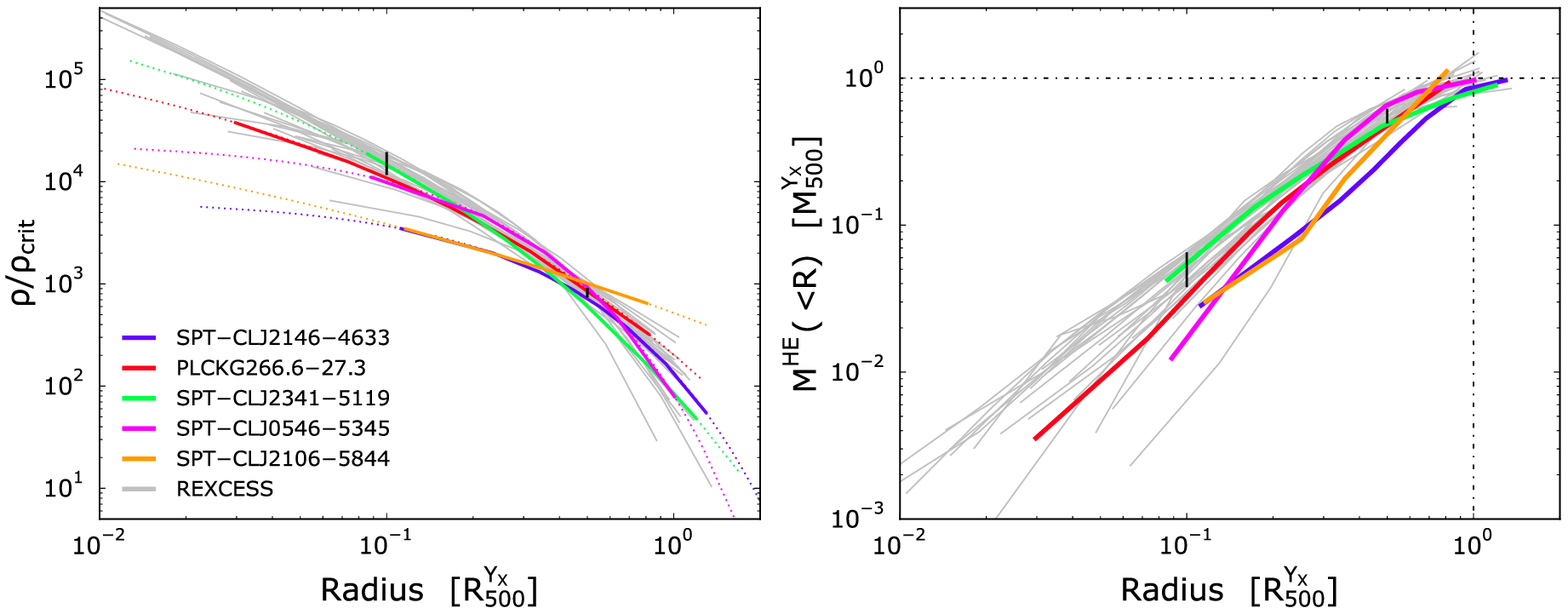} 
\end{center}
\caption{\footnotesize{\textit{Left panel}: Scaled total density profiles computed using the mass profiled derived from the DF Einasto model.  For each cluster, the total radial range is that of the combined density profile; estimates beyond the radial range covered by temperature measurements are marked with dotted lines. 
The grey lines represent the scaled total density profiles derived from the \rexcess\ sample. \textit{Right panel}: scaled mass profiles. The colour scheme is the same as in the left panel. The black error bars in both panels represent the 68\% dispersion of the \rexcess\ profiles at $0.1$ and $0.5$ $R_{500}^{Yx}$.}}
\label{fig:total_rho}
\end{figure*}

\section{Evolution of cluster properties}\label{sec:evolution}

\subsection{$\Mvyx$--$\Mvhe$ relation and evolution of the ratio}

The \MY\ relation we use in this work was calibrated using  hydrostatic masses derived from the relaxed subsample of 12 \rexcess\ objects \citep{arnaud2010,pra10}, plus eight additional relaxed systems from \citet{arn07}. This relation was derived from local objects and we assumed  self-similar evolution. The present observations offer the opportunity to investigate the robustness of this relation when applied to a high-redshift sample dominated by disturbed objects. Figure \ref{fig:yx_vshy} shows the resulting comparison of  $\Mvyx$ with  $\Mvhe$ and $\Mvheyx$, in the left and right panels, respectively.  

In both cases there is excellent agreement between individual measurements. The only exception is the $\Mvhe$ of \five, which is subject to  the systematic uncertainty discussed above. 
The error-weighed mean ratios are  $\Mvyx/\Mvhe=1.02^{+0.15}_{-0.13}$ and $\Mvyx/\Mvheyx=1.04^{+0.09}_{-0.08}$, consistent with unity.
This suggets that the relation is robust, even when applied to such an extreme sample. The good agreement is consistent either with no evolution of the ratio between the two quantities, or with an evolution of the ratio where the evolution is counterbalanced by some other effects. However the latter explanation is unlikely given that the evolution is perfectly compensated, such that the agreement between $\Mvyx$ and $\Mvhe$ is excellent as a function of redshift. We note that this result does not necessarily imply that there is no evolution of the bias between the hydrostatic mass and the true mass. However, our comparison with weak lensing above would suggest that the bias cannot be dramatic.
%


\subsection{Scaled mass and total density profiles}\label{scaled_masses}

We calculated the total density profiles for our sample using the best-fitting DF Einasto model. The resulting profiles are shown compared to those from \rexcess\ in the left panel of \figiac{fig:total_rho}. The right panel shows the corresponding cumulative total mass profiles. 

A consistent picture emerges from these comparisons. At $\Rvyx$ all the mass profiles are in excellent agreement: the dispersion is similar, and interestingly is centred around unity (i.e. the  $\Mvhe$ is comparable to $\Mvyx$, consistent with our findings in the previous section).  Apart from the profile of \five, which is affected by poor radial coverage especially at large radius, all the density profiles of the $z\sim 1$  sample lie within the envelope of the \rexcess\ profiles at high radii ($> 0.5\Rv$). However, the profiles tend to be shallower on average in the central regions. The profiles of SPT-CLJ2146-4633 and \five\ are  even shallower in the core ($<0.3\Rv$) than the least-peaked \rexcess\ profile.  The other three systems lie within the $1\sigma$ dispersion of \rexcess\ profiles, but  tend to trace the lower envelope of the distribution in total density and total mass, especially towards the most central parts.

Unfortunately, due to the small size of the sample and the poor quality of \five, we cannot quantify whether there is a significant difference compared to \rexcess in median mass profile shape and/or an increase in the intrinsic scatter around it. If these differences are confirmed, this behaviour can be interpreted either as evolution in the core regions or as being due to a difference between X-ray and SZ selection. Comparison with an X-ray selected sample at similar redshifts or comparison to a similar SZ-selected sample at lower redshift, would help to clarify this point.

\begin{figure}[!t]
\begin{center}
\includegraphics[width=0.8\columnwidth]{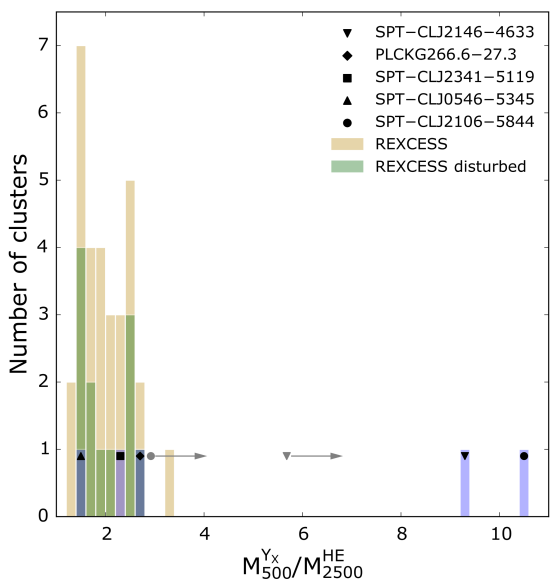} 
\end{center}
\caption{\footnotesize{Number of clusters as a function of their sparsity. The blue and gold shaded bins represent the sparsity distribution of the five high-$z$ clusters and the \rexcess\ sample, respectively. Individual objects are identified by symbols  over-plotted on the blue bins. The grey arrows represent the lower limit of the sparcity of \one\ and \five.}}
\label{fig:spar}
\end{figure}



\subsection{Sparsity}

The halo sparsity was introduced by \cite{balmes2014} to characterise the form of the mass distribution in a way that is independent of any parametric model. It is defined as the ratio of masses integrated within two fixed overdensities,
\begin{equation}
S \equiv \frac{M_{\Delta_{1}}}{M_{\Delta_{2}}},
\end{equation}
where $\Delta_{1,2}$ represent the overdensities at which the masses are calculated, with $\Delta{1}<\Delta{2}$.  As is discussed in \cite{balmes2014}, the properties of the sparsity are independent of the choice of the $\Delta$ as long as the definition of the halo is not ambiguous ($\Delta_{1}$ not too small), and that dynamical interaction between baryons and dark matter  can be neglected ($\Delta_{2}$ not too large).  

We chose to measure the sparsity within overdensity of $\Delta_{1} = 500$ and $\Delta_2 = 2500$ with respect to the critical density. These overdensities are well matched to the sensitivity of the X-ray observations discussed here, and are sufficiently distant to properly sample the form of the mass profile.

 Figure \ref{fig:spar} shows the resulting sparsity measurements for our $z\sim1$ sample. These data are compared to those from \rexcess, which exhibit a peaked distribution in a narrow range, $1<S<3$. 
 
Three of the clusters in our $z\sim1$ sample have sparsity values that lie well within the \rexcess\ distribution. \one\ and \five\ lie outside this distribution, their sparsity being  $\sim 4-5$ times the mean value ($\sim 2$) compared to \rexcess. This result reflects what we already found for the mass profiles. This study and the recent parallel study of  \citet{cor17} represent the  first applications of this quantity to a large sample of objects. The narrow distributions in \figiac{fig:spar} show its  effectiveness in tracing the population characteristics. 

\begin{figure}[!t]
\begin{center}
\includegraphics[width=0.45\textwidth]{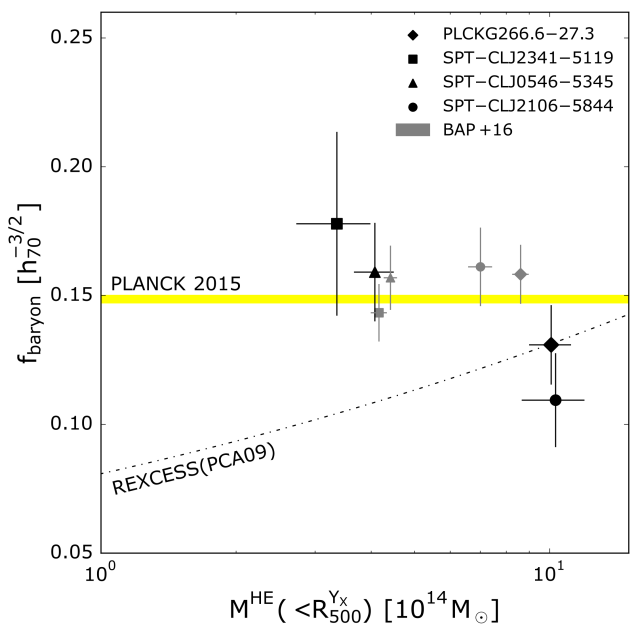} 
\end{center}
\caption{\footnotesize{Baryon fractions computed at $\Rvyx$ as a function of mass. The baryon fraction does not show any dependence with respect to the mass at this high $z$. Black points represent the baryon fraction computed using the hydrostatic mass profiles at $\Rvyx$. The grey points represent the baryon fractions published in \cite{bartalucci2016}, computed using the $\Mvyx$. Gas masses were computed using the gas mass profiles derived from the combined density profiles. We used the stellar masses published in \cite{chiu2016}; stellar mass for \one\ is not available. The yellow shaded area represents the baryon fraction published in \cite{planck2015}. }}
\label{fig:gas_frac}
\end{figure}

\subsection{Baryon fraction}

The baryon fraction determined at the radius $R$ is defined as
\begin{equation}\label{eq:bar_frac}
f_{baryon}=(M_\textrm{star}+M_\textrm{gas})/M_\textrm{tot} ,
\end{equation}
 where $M_\textrm{star}$ is the total stellar mass, $M_\textrm{gas}$ is the gas mass, $M_\textrm{tot}$ is the halo total mass, and all quantities are integrated within $R$. In B17 we presented the baryon fraction derived using $\Mvyx$ for the total mass estimate. Here we extended this analysis by deriving the baryon fraction using $\Mvheyx$  for the $M_\textrm{tot}$ term. This is fundamental to understand possible systematics related to the fact that the gas mass profiles and $\Mvyx$ measurements are correlated (i.e. the $\YX$ is based on the gas mass). Figure \ref{fig:gas_frac} shows the baryon fraction as a function of mass computed for this work, the results from B17, and the mean derived from \rexcess\ \citep{pratt2009}. For $M_\textrm{star}$ at $R_{500}$ we used the stellar masses published in \citet[][the stellar mass for \one\ is not available]{chiu2016}. 

The baryon fractions for \two\, \three\, and \four\ are in excellent agreement with the previous results published  in B17. \five\ presents a larger deviation, but  the hydrostatic mass computation for this object is affected by the lack of radial coverage. The use of hydrostatic mass measurements here confirms and consolidates what we found in B17: in this redshift regime, the baryon fraction does not show any dependence with respect to the mass. The density enclosed within a certain radius is higher hence more energy is required to expel the gas.  We also confirm the good agreement between the baryon fraction of our sample with the fraction derived by the Planck collaboration \citep{planck2015}. 
\begin{figure}[!t]
\begin{center}
\includegraphics[width=0.45\textwidth]{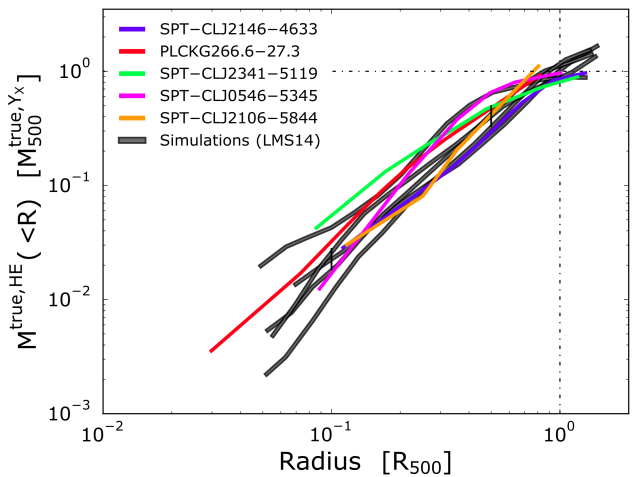} 
\end{center}
\caption{\footnotesize{Scaled hydrostatic mass profiles derived in this work and from the suite of cosmological simulations published in \cite{amandine2014}, shown with coloured and grey solid lines. The black error bars represent the 68\% dispersion of the simulated profiles computed at $0.1$ and $0.5$ $R_{500}^\mathrm{true}$.
Simulated and X-ray mass profiles were scaled by their $M_{500}^\mathrm{true}$ and $\Mvyx$, respectively. Our sample and simulated cluster radial profiles were scaled by their $\Rvyx$ and $R_{500}^\mathrm{true}$, respectively. The common scaled radius is indicated with $R_{500}$.}}
\label{fig:simulation_comp}
\end{figure}
\section{Comparison with simulations}\label{sec:simulations}

\subsection{Mass profiles}
We now turn to a comparison with cosmological numerical simulations. We use the same simulated sample of five  $z=1$ galaxy clusters in the $[4-6]\times 10^{14} \times M_{\odot}$ mass range described in Sect.~6 of B17, selected from the AGN 8.0 model of the suite of hydrodynamical cosmological simulations cosmo-OWLS \citep{amandine2014}. These simulations include baryonic physics, and represent an extension to larger volumes of the OverWhelmingly Large Simulations project (\citealt{schaeye2010}). 

From the simulated datasets we extracted and fitted the pressure profiles using a generalised NFW model. We then derived the simulated mass profiles by applying the hydrostatic assumption to the gNFW pressure profile in combination with the density profile \citep[see e.g.][]{pra16}. Figure \ref{fig:simulation_comp} shows the comparison between the observed FNPL and simulated mass profiles, scaled by $\Mvyx$ and $M_{500}^\textrm{true}$, respectively, where $M_{500}^\textrm{true}$ is defined as the sum of all the particles within  $R_{500}^\textrm{true}$. 

The agreement over the full radial range is remarkably good. The shape, normalisation, and scatter of the simulated profiles seem to reproduce well the observations, four of the five observed profiles lie within the $68\%$ dispersion of the theoretical profiles computed at $0.1$ and $0.5$ $R_{500}^\mathrm{true}$. 
Interestingly, there is also excellent agreement of the profiles at $\Rv$, hinting that $\Mvyx$ represents a robust estimate of the true mass in this mass and redshift regime. Unfortunately,  we were not able to investigate the behaviour of the profiles  in the core regions below $0.1\Rv$. Furthermore,  as the five simulated clusters discussed here are the only objects in the cosmo-OWLS cosmological box that fulfil the mass and redshift criteria, the qualitative agreement might be coincidental.
A larger number of higher resolution simulations and better sampling of the X-ray profiles are needed in order to make progress on this front.

\begin{figure}[!t]
\begin{center}
\includegraphics[width=0.8\columnwidth]{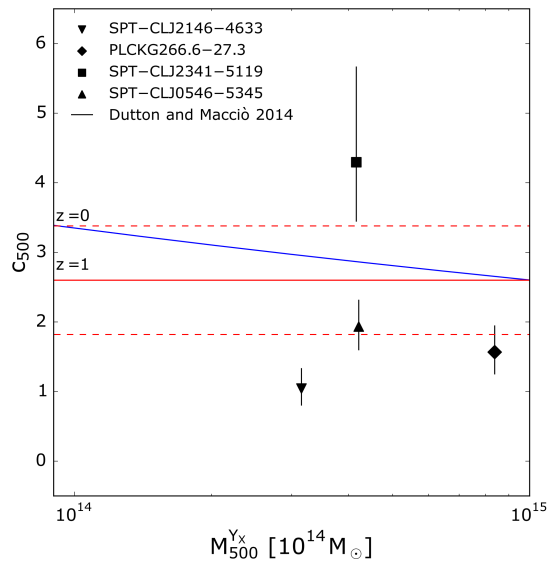} 
\end{center}
\caption{\footnotesize{\textit{Right panel}: Concentration-mass relation. The $c_{500}$ is derived from the DF NFW model. Blue and red solid lines represent the theoretical relations from the suite of cosmological simulations of \cite{dutton2014}. Dotted lines represent the $30\%$ scatter \citep{bhatt2013} for the $z=1$ relation. The concentration of \five\ is not reported because the data radial coverage does not allow a robust determination of $c_{500}$.}}
\label{fig:c500}
\end{figure}

\subsection{Concentration}

The NFW concentration is known to evolve with redshift and mass \citep[e.g.][]{dutton2014}, although at the highest masses there is surprisingly little evolution \citep{leb18}. While the mass dependence in the local Universe has been confirmed in a number of works \citep[e.g.][]{pra05,pap05,vikhlinin2006,voi06,gas07,buo07,ett10}, evolution has received less attention \citep{ser13,sch08,amodeo2016}. The constraints are especially poor in the high-$z$ regime;  the  typical uncertainties on concentration parameters of the  five clusters at $z>0.9$  studied by \citet{amodeo2016} are of the order of  $\pm[60$--$80]\%$ (their Table 2).

The very precise measurements afforded by the present observations allow us to further investigate the $c-M$ relation and its evolution. Figure \ref{fig:c500} shows the concentrations for four of the clusters in our sample compared to the theoretical predictions derived from the simulations of \cite{dutton2014}. The predictions were computed for a set of clusters in the local and distant universe, at $z=0$ and $z=1$, respectively, at $\Delta=200$. We translated their results at $\Delta=500$ using the NFW profile.  At  $z=1$, we considered the concentrations plotted in their Fig. 10 rather than  their power law fit, the c--M relation flattening to $c_{200}=4.$ ($c_{500}=2.6$) in the present high-mass range. \cite{bhatt2013} found that the dispersion for the $c-M$ relation is $\sim 30\%$. We used this result to roughly estimate the typical dispersion for the  $c-M$  at $z=1$.   

Two clusters are within the $1\sigma$ dispersion of the mean expected relation, while two are $[1.5$--$2]\sigma$ away. We iteratively computed the mean concentration, taking into account statistical errors and intrinsic scatter.  Our results agrees, at the $1\sigma$ level, with the expectations: the mean concentration is $\left<\,c_{500}\,\right>\,=\,2.06\pm0.67$, with an estimated intrinsic dispersion of $1.2\pm0.6$. Although the sample size is small, this is the first  test of the $c$--$M$ relation at these redshifts with precise  individual concentration measurements (i.e. errors  smaller than the expected scatter).

\section{Discussion and conclusions}\label{sec:conclusions}

We have presented the {\it individual} hydrostatic mass profiles of the five most distant ($z\sim 1$) and massive ($\Mvsz>5\times 10^{14}$) galaxy clusters from the SPT and Planck cluster catalogues, measuring for the first time the profiles up to $R_{500}$.
 The combination of \chandra\ and \xmm, following the technique developed in B17, allowed us to overcome cosmological dimming and to derive robust measurements from the core regions out to $\Rv$. 
The temperature profiles cover a typical radial range of $[0.08 - 1]\,\Rv$, while the combined \xmm/\chandra\ density profiles are typically in the range $[0.01 -1.7]\,\Rv$. We considered both parametric (forward and backward) and non-parametric approaches to measuring the mass profiles. The main results regarding the robustness of the X-ray profiles are the following:
\begin{itemize}
\item X-ray hydrostatic mass measurements at this redshift regime are remarkably robust and method-independent. All the profiles are consistent within the uncertainties as long as they are determined in the radial range where there are density and temperature measurements. This robustness is also reflected in the determination of mass at fixed radius or at a particular density contrast. 
\item In the very core region $R<0.08\,\Rv$, where only density information is available, parametric models are necessary. The density information brings a certain constraint to the shape of the mass profile, but with an uncertainty that increases with decreasing radius.  
\item At $\Rv$, it is essential to have a temperature measurement to anchor the total mass at this radius and constrain the shape of the mass profile. Robust $\Mv$ estimates are only possible when this condition is fulfilled. In the absence of this constraint, model extrapolation can be rapidly divergent and can yield unphysical results.
\item Generally, when the radial sampling is poor (the profiles have few points) or when the profile is irregular, estimation of the mass outside the radial range probed by the temperature data is less robust and will depend strongly on the method used to measure the mass. On the other hand, if the shape of the profile is well reproduced using an NFW or Einasto-type model, the resulting mass estimate outside the range with measured temperatures is more robust. 
\item We compared $\Mvhe$ and $\Mvyx$ for four clusters of our sample with weak lensing (WL) mass measurements from {\it HST} observations, finding that the X-ray and WL mass measurements are in agreement within the uncertainties. 
There is, however, an offset on average, in the sense that the X-ray masses appear to be systematically higher  by a factor of  $1.39^{+0.51}_{-0.37}$ than the WL masses. 
This offset goes in the opposite direction to what has been found in previous works (e.g. WtG at the $2\sigma$ level), and is contrary to the expectations for a `hydrostatic bias'. 
\end{itemize}

The above results confirm the power of combining \xmm\ and \chandra\ for measuring the mass profile distribution and for estimating the hydrostatic $\Mv$ up to $z\sim1$. We expect these results to be even less sensitive to systematic effects such as background estimation, contamination by background or foreground point sources, and the absolute temperature calibration than for local clusters. This is because object angular size is much smaller than the field of view, yielding a better constraint on the background, and the spectrum is redshifted to lower energies, where the effective area calibration of X-ray telescopes is more robust. In parallel, the \chandra\ observations allow robust point source detection and density measurement very deep into the core regions. In contrast, WL measurements become increasingly challenging at these redshifts. The statistical quality of the WL mass data is much poorer than is reachable with X-rays, even with {\it HST}, and control of systematic effects (in particular the measure of the redshift distribution of background sources or the removal of contamination by cluster members) becomes more demanding. The fact that we find a positive HE bias is probably linked to these effects. \\

We then investigated the evolution by comparison with local data and with expectations from numerical simulations. The main results were:
\begin{itemize}

\item The agreement between the hydrostatic masses at $\Rvyx$ or at $\Delta=500$ and $\Mvyx$ is remarkably good, suggesting that the \MY\ relation is robust and that it can be extended to samples of disturbed and distant objects. It also suggests that there is no significant evolution between $\Mvyx$ and $\Mvhe$ with redshift. The comparison with WL masses would further suggest that there is no dramatic increase in the bias between the hydrostatic mass and the true mass. However, it is clear that better WL data are needed to settle this point. 
\item We compared the scaled mass and total density profiles to those of the X-ray selected local sample \rexcess. This comparison shows that on average there is excellent agreement with \rexcess\ at large radii. The clusters of our sample exhibit a larger dispersion in shape over the full radial range, and systematically trace the lower envelope of the \rexcess\ distribution in the core region. These results suggest either the presence of evolution, or an X-ray / SZ selection effect. 
\item We computed the sparsity  for a large sample of clusters  (the five high-$z$ objects plus \rexcess) studied with X-ray observations. The sparsity enables efficient characterisation of the mass distribution in the cluster halo, and the comparison with \rexcess\ confirms the above. 
\item We extended and strengthened the baryon fraction results found in B17. Using the hydrostatic mass measurements we confirmed our previous finding indicating that the baryon fraction at this redshift does not depend significantly on the halo mass, and agrees with the value from \cite{planck2015}.
\item A comparison with the cosmo-OWLS simulations \citep{amandine2014} showed that there is excellent agreement between observed and simulated profiles, the latter derived by imitating an X-ray approach. The scatter of our sample is also well reproduced by the simulations over the full radial range. We also studied the concentration-mass relation for the first time at high precision in this mass and redshift range, and found  good agreement with the evolution predicted by \cite{dutton2014}.

\end{itemize}

This work represents the first full application of the method developed in B17, confirming that the combination of \chandra\ and \xmm\ is crucial in order to study high-redshift objects, and allowing us to investigate the statistical properties of the mass profiles of cluster haloes in the high-mass, high-redshift $z\sim1$ range. Despite the small sample size, we were able to obtain a first insight into the statistical properties of these cluster haloes, suggesting profiles that are slightly less peaked than in  local systems, in line with the expected theoretical evolution. However, a robust low-redshift SZ-selected anchor for the radial mass distribution is badly needed, especially taking into account the now well-known issue of X-ray versus SZ selection effects \citep{lov17,and17,ros17}. Larger sample sizes are needed to better consolidate the average behaviour {\it and} its dispersion. In parallel, higher resolution numerical simulations of larger volumes \citep[e.g.][]{leb18} are needed to provide the theoretical counterparts to the type of objects we have studied here.

\begin{acknowledgements} 
The results reported in this article are based on data obtained from the \chandra\ Data Archive and observations obtained with \xmm, an ESA science mission with instruments and contributions directly funded by ESA Member States and NASA. This work was supported by CNES. The research leading to these results has received funding from the European Research  Council  under  the  European  Union’s  Seventh  Framework
Programme (FP72007-2013) ERC grant agreement no 340519. 

 \end{acknowledgements}

\bibliographystyle{aa}
\bibliography{lib_articoli}

\appendix
\section{Parametric models used}\label{sec:parametric_models}
In this section we report the parametric models based on \cite{vikhlinin2006} we used in this work. 
We fitted the combined density profiles with
\begin{equation}\label{eq:dens}
\begin{split}
n_{e}(r) =  n_{01} \frac{ (r/r_{c1})^{-\alpha} }{(1 + r^2/r_{c1}^{2})^{3\beta_{1}/2 - \alpha/2}} & \frac{1}{(1+r^3 / r_{s}^3)^{\epsilon/3}} + \\ & +  \frac{n_{02}}{(1 + r^2/r_{c2}^2)^{3\beta_{2}/2}},
\end{split}
\end{equation}
where $n_{01}$, $r_{c1}$, $\alpha$, $\beta_1$, $r_{s}$, $\epsilon$, $n_{02}$, $r_{c2}$, and $\beta_2$ were free parameters.
Deprojected 3D temperature profiles were fitted with:
\begin{equation}\label{eq:temp}
T_{3D} (r) = T_{0} \frac{ (r/r_{cl})^{a_{cl}} + \tau}{  (r/r_{cl})^{a_{cl}} + 1} \frac{1 }{ \left[ 1 + (r/r_{t})^b \right]^{c/b}, }
\end{equation}
where $T_0$, $T_{min}$, $r_t$, $b$, $c$, and $r_{cool}$ were free parameters. We fitted the temperature profile of \four\ fixing both $a_{cl}$ and $b$ to $2$. The temperature profile of \five\ was fit using a third-degree polynomial.

\section{Density profile of SPT-CLJ0546-5345}\label{sec:appendix_spt_ne}
\begin{figure}[!ht]
\begin{center}
\includegraphics[width=0.45\textwidth]{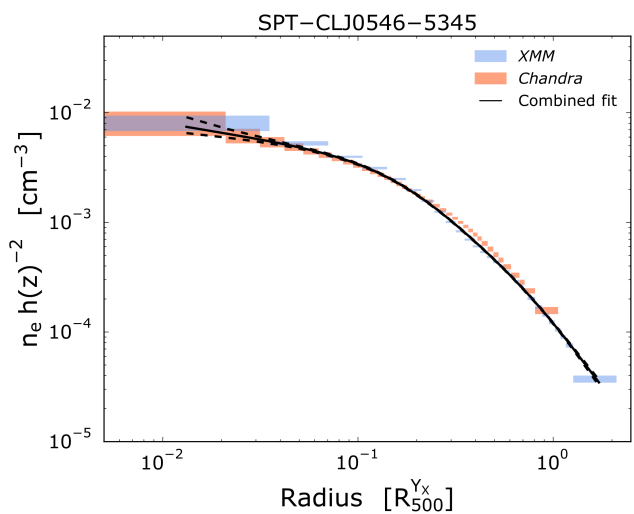} 
\end{center}
\caption{\footnotesize{Normalised, scaled, and deprojected density profile of SPT-CLJ0546-5345 measured by \chandra\ and \xmm\ with red and blue polygons, respectively. 
The black solid line and the dotted lines represent the simultaneous fit with the density parametric model of \citealt{vikhlinin2006} and its $1\sigma$ error, respectively.}}
\label{fig:spt5345ne}
\end{figure}

This work is focused on the extraction of mass profiles under the assumption of HE. For this reason, we masked the substructure in the south-west sector of \four, highlighted with the blue dotted circle in Fig. A.1 in B17, and derived the density and temperature profiles centred on the X-ray peak. The details of profiles extraction are given in B17.
Figure \ref{fig:spt5345ne} shows the deprojected density profiles of \four\ using \chandra\ and \xmm\ datasets. Given the excellent agreement between the two, the profiles were simultaneously fitted using the parametric model of \citealt{vikhlinin2006}. Its uncertainties were calculated using a Monte Carlo procedure. 

\section{Mass profile errors}\label{sec:appendix_mass}

The relative errors of the hydrostatic mass profiles derived using the NFPL and the FP method shown in \figiac{fig:mass_profiles} exhibit a radial dependency. In the core the relative errors are larger than in the outskirts of the profiles. This is counter-intuitive because the mass profiles were derived from the density profiles for which the relative error is negligible ($\sim 1-2\%$) and from temperature profiles which were defined to have a the same signal-to-noise ratio in each radial bin (see Section 3.4 of B17).
The relative error of the mass profile $M(<r)$ derived using Eq. \ref{eq:mass} and neglecting the error on the density profile is:
\begin{equation}\label{eq:appendix}
\frac{\Delta M}{M} \propto \sqrt{\left(\frac{\Delta T}{T} \right)^2 + \left(\frac{\Delta \beta_{T}(R)}{\beta_{T}(R) + \beta_{n_e}(R)} \right)^2},
\end{equation}
where $\beta$ is defined as the logarithmic derivative of temperature and density, namely $\beta_T(R)$ and $\beta_{n_e}(R)$, respectively, with respect to the logarithmic derivative of the radius, $\beta \equiv \dif \log{x}/\dif\log{r}$. Equation \ref{eq:appendix} shows that the relative error is proportional to the sum of the term $\mathrm{A} = \Delta T/ T$, and of the term $\mathrm{B}= \Delta \beta_{T} / (\beta_{T} + \beta_{n_e})$. 

\begin{figure*}[!ht]
\begin{center}
\includegraphics[width=1\textwidth]{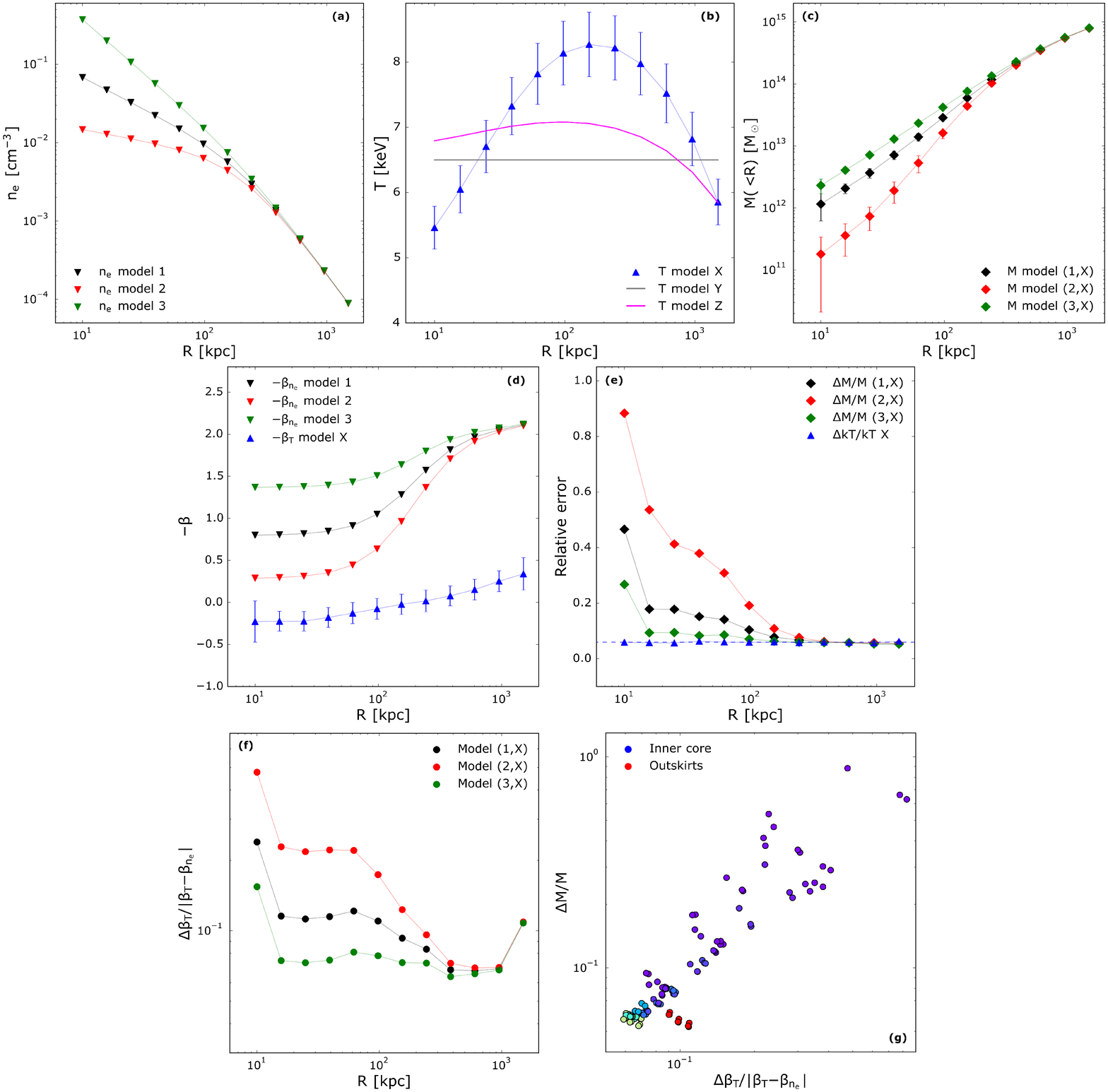} 
\end{center}
\caption{\footnotesize{For all the panels except panel \textit{g}: points represent the quantity measured at fixed radial bins. The solid line is showed to guide the eye. 
\textit{Panel a:} density profiles generated using the parametric model of \cite{vikhlinin2006}. The errors are set to 0. 
\textit{Panel b:} temperature profiles generated using the parametric model of \cite{vikhlinin2006}. The relative error on each bin is fixed to 6\%. For clarity, bins and corresponding errors are reported only for `T model X'. We report only the shape form of `T model Y' and `T model Z' with solid grey and magenta lines, respectively. 
\textit{Panel c:} hydrostatic mass profiles derived from the three density profiles and from the temperature profile `T model X' shown in panel \textit{a} and \textit{b}, respectively, using the FP method. 
\textit{Panel d:} $\beta$ is defined as the logarithmic derivative of the density and temperature, namely $\beta(n_e)$ and $\beta(kT)$,  with respect to the logarithm of the radius, $\beta \equiv \partial \ln{x}/ \partial \ln{r}$. The errors of $\beta(kT)$ are derived via a Monte-Carlo procedure.
\textit{Panel e:} temperature and mass profile relative errors as a function of the radius. The dotted line represents the 6\% relative error. The mass profiles are computed using the three density profiles and the `T model X'.
\textit{Panel f:} ratio between the error of $\beta(kT)$ and the distance $D$ between $\beta(n_e)$ and $\beta(kT)$, the distance being defined as 
$D \equiv |\beta(kT) - \beta(n_e)$. 
\textit{Panel g:} mass profile relative errors as a function of the same quantity shown on the $y$ axis of panel \textit{f}. For this plot, we also included the results using all the temperature profiles. Points are colour-coded in order to clearly identify the inner core (blue points, upper right part of the plot) and the outskirts (red points, lower left part of the plot).}}
\label{fig:mass_errors}
\end{figure*}

The behaviour of the two terms as a function of the radius can be studied using simple models for the density and temperature. 
We employed the density and temperature parametric models in Eq. \ref{eq:dens} and Eq. \ref{eq:temp}, respectively, to generate `toy model' temperature and density profiles. The parameters are reported in Table \ref{tab:toy_stuff}. 
We neglected the error on the density profiles and assumed a constant relative error on the temperature profile of 6\%. These assumptions are a good approximation of a realistic case, where density profiles are well constrained and temperature profiles are tailored to have 
constant signal-to-noise ratio. 
We generated three density profiles with different inner slopes and three temperature profiles with different shapes, shown in panels \textit{a} and \textit{b} of \figiac{fig:mass_errors}, respectively. 
These profiles are representative of what is generally found in large samples of clusters. The density profile in the core strongly varies from cluster to cluster (cool-core, dynamically disturbed, relaxed, etc.). In the outskirts the behaviour is self-similar and all the profiles present a steep gradient. 
The temperature profile strongly depends on the cluster characteristics. The shape ranges from  the `bell' shape of the cool-core clusters (T model X) to
being almost flat (T model Z). For a gallery of individual density and temperature profiles, see e.g. \cite{vikhlinin2006}, \cite{pratt2007}, and \cite{croston2008}.

We took as reference the temperature profile `T model X', and computed the hydrostatic mass profiles for the three density toy models. The results are shown in panel \textit{c} of \figiac{fig:mass_errors}.
We observe the following:
\begin{itemize}
\item the logarithmic slope, $\beta$, of the density and temperature profiles strongly depends on the radius. Panel \textit{d} shows that $\beta_T$ is small at all radii and smaller than $\beta_{n_e}$. The difference between the two increases with radius;
\item the relative error of the mass profile $\Delta M / M$, shown in panel \textit{e}, reflects this different behaviour of the $\beta$s.In the outskirts, $\beta_{n_e}$ is much larger than $\beta_{T}$ so that the B-term in Eq. \ref{eq:appendix} becomes negligible compared to the A-term. For this reason, the mass profile relative errors in the outskirts tend to coincide with the temperature relative error, i.e. the A-term. This is not the case in the core where the difference between $\beta_T$ and $\beta_{n_e}$ is less important and the B-term is no longer negligible. For this reason, in the core the mass relative errors are larger than the A-term only. The relative importance of the B-term is related to the distance between the two $\beta$s;
\item the behaviour of the B-term as a function of radius can be visualised studying the ratio between $\Delta\beta_T$ and the distance $D$ between the $\beta_{n_e}$ and $\beta_T$, the distance being defined as $D \equiv |\beta_T - \beta_{n_e}|$. Panel \textit{f} in \figiac{fig:mass_errors} shows that this ratio decreases with radius as $D$ increases. The mass measurements in the final radial bin slightly deviate from this behaviour because of the larger error of the $\beta_T$ in the last bin. We derived $\beta_T$ and its error $\Delta\beta_T$ by estimating the median and the $68\%$ deviation within $1000$ realisations of the temperature profile. For each realisation, we estimated the gradient for the n-th bin determining the slope using the $[n-1,n,n+1]$ bins. For the boundary bins we determined the slope selecting the first and last three radial bins, respectively. This is less constraining for the gradient so that the dispersion within all the realisations is greater and the resulting error $\Delta\beta_T$ is larger;
\item there is a clear correlation in the log-log space between the relative error on the mass $\Delta M/M$ and this ratio $\Delta\beta_T/D$, as seen in panel \textit{g}. Results using the other two temperature profiles are also shown and present the same behaviour.  
\end{itemize}
The behaviour of the relative error on the mass profile is thus an intrinsic property of the hydrostatic equation, Eq. \ref{eq:mass}, and does not depend on the temperature or density profile shape. In particular, this effect is tightly linked to the general behaviour of the temperature and density gradients in galaxy clusters.

\begin{table}[!h]
\caption{{\footnotesize Parameters of the toy models.}}\label{tab:toy_stuff}
\begin{center}
\resizebox{0.9\columnwidth}{!} {
\begin{tabular}{cc}
\hline        
\hline
Model  					        		      & Parameters \\
											  &					\\
$n_e$ model $1^{a}$      & $n_{01}=6.28 \times 10^{-3}$ cm$^{-3}$, $r_{c1}=198$ kpc, $\beta_1=0.7$,  \\
											&	$\alpha=0.57$, $r_s=1613$ kpc, $\epsilon=0.1$,        \\
											& $n_{02}=10^{-3}$ cm$^{-3}$, $\beta_2=0.67$, $r_{c2}=19.8$ kpc \\
											& \\
$T$ model X$^{b}$     	&  $T_0=9.1$ keV, $\tau=0.46$, $r_{cl}=22.78$ kpc, \\
											& $a_{cl}=1.21$, $r_t=7391$ kpc, $b=1.03$, $c=2.56$ \\
						
\hline

\end{tabular}
}
\end{center}
\footnotesize{Notes: $^{(a)}$ $n_e$ models 2 and 3 were obtained using the same parameters as that of model 1 except for $\alpha = 0.29 $ and $\alpha= 1.37$, respectively. $^{(b)}$ model Y was obtained using the same parameters as model X except for $T_{0}=7.28$, $\tau=0.91$, and $c=1.28$. The temperature model Z is a flat profile with $T=6.5$.}
\end{table}

\end{document}